\documentclass{article}
\usepackage{graphicx}  
\usepackage{amsmath}   
\usepackage[compress]{cite}
\usepackage{amssymb}   
\usepackage{bm} 
\usepackage{dcolumn}
\usepackage{color}
\usepackage{mathrsfs}
\usepackage{amsfonts}
\usepackage{varioref}
\usepackage{braket}
\usepackage{subcaption}
\RequirePackage[colorlinks,citecolor=blue,urlcolor=magenta,linkcolor=blue]{hyperref}
\def\eq#1{{Eq.~(\ref{#1})}}

\def\frab#1#2{\left(\frac{#1}{#2}\right)}

\def\ket#1{|#1\rangle}                    
\def\bk#1#2#3{{\langle #1|#2|#3\rangle}}  
\def\amp#1#2{\langle #1 | #2\rangle}      

\title{Thermality of the Rindler horizon: A  simple derivation from the structure of the inertial propagator}
\author{T. Padmanabhan\\
IUCAA, Pune University Campus,\\
Ganeshkhind, Pune 411007, India.\\
Email: paddy@iucaa.in}
\date{ }  

\begin{document}
	\maketitle
	
	\begin{abstract}
	The Feynman propagator $G(x_1,x_2)$ encodes \textit{all} the physics contained in a free  field and transforms as a covariant biscalar. Therefore, we should be able to discover the thermality of Rindler horizon, just by probing the structure of the propagator, expressed in the Rindler coordinates. I show that the thermal nature of the Rindler horizon is indeed contained --- though hidden --- in the standard, inertial, Feynman propagator. 
	The probability $P(E)$ for a particle to propagate between two events, with energy $E$, can be related to the temporal Fourier transform of the propagator. A strikingly  simple computation reveals that: (i) $P(E)$ is equal to $P(-E)$ if the propagation is between two events in the same Rindler wedge while (ii) they are related by a Boltzmann factor with temperature $T=g/2\pi$, if the two events are separated by a horizon. A more detailed computation reveals that the propagator itself can be expressed as a sum of two terms, governing  absorption and emission, weighted correctly by the factors $(1+n_\nu)$ and $n_\nu$ where $n_\nu$ is a Planck distribution at the temperature $T=g/2\pi$. In fact, one can \textit{discover} the Rindler vacuum and the alternative (Rindler) quantization, just by probing the structure of the inertial propagator.  These results can be extended to local Rindler horizons around any event in a curved spacetime. The implications are discussed.  
	\end{abstract}

\section{The main result: Inertial propagator knows all!}
 The path integral representation of the (Feynman) propagator is given by the sum over paths prescription using the (square-root) action for a relativistic particle:
 \begin{equation}
\sum_{\rm paths} \, \exp\left[ - i m\ell(x_1,x_2) \right] = G(x_1,x_2)
  \label{rt0}
 \end{equation} 
  where $\ell (x_1,x_2)$ is the length of the path. This suggests that one can interpret $G(x_1,x_2)$  as an amplitude for a particle/antiparticle to propagate between two events in the spacetime.\footnote{I use mostly negative signature --- except when specified otherwise --- and natural units. The propagator in momentum space $G(p)=i(p^2-m^2-i\epsilon)^{-1}$ is defined with an $i$ factor, so that  $G(x_1,x_2)=\bk{0}{T[\phi(x_1)\phi(x_2)]}{0}$.}. This interpretation acquires an operational meaning in the presence of a source $J(x)$ capable of emitting/absorbing the particles \cite{tpqft}. Then the vacuum persistence amplitude
  \begin{equation}
\amp{\text{out}}{\text{in}}_J =\amp{\text{out}}{\text{in}}_{J=0} \exp\left\{ - \frac{1}{2} \int d^Dx_1\, \sqrt{-g_1}\ \int d^D x_2 \sqrt{-g_2} \ J(x_1)\, G(x_1, x_2)\, J(x_2)\right\}
   \label{rts1}
  \end{equation}
  can be thought of describing the emission/absorption at the two events (controlled by $J(x_1),J(x_2)$) and the propagation between the events governed by $G(x_1,x_2)$.
  
  I am interested in the stationary situations in which the propagator depends  on the time coordinates only through the time difference,  so that $G(x_1,x_2) = G(\tau; \bm{x}_1,\bm{x}_2)$ with $\tau \equiv (x_1^0-x_2^0)\equiv (\tau_1 - \tau_2)$. Such  stationarity is assured  if there exists a Killing vector field $\xi_a$, which, in suitable coordinate system, can be represented as $\xi_a = \partial/\partial \tau$. 
  One can then interpret the temporal Fourier transform
  \begin{equation}
   A(\Omega; \bm{x}_1,\bm{x}_2) = \int_{-\infty}^\infty d\tau\ G(\tau; \bm{x}_1,\bm{x}_2) \ e^{i\Omega \tau};
   \qquad \tau=(\tau_1-\tau_2)
   \label{rt1}
  \end{equation} 
  as the amplitude for the particle to propagate between $\bm{x}_1$ and $\bm{x}_2$ with energy $\Omega$, introduced  as the Fourier conjugate to the time coordinate $\tau$. In what follows I will simplify the notation and write  $G(\tau) $ for $G(\tau; \bm{x}_1,\bm{x}_2)$ and $A(\Omega)$ for $A(\Omega; \bm{x}_1,\bm{x}_2)$, suppressing the spatial coordinates. While evaluating the amplitude $A(\Omega)$ in \eq{rt1} it is convenient to  assume that $\Omega > 0$ and  interpret $A(-\Omega)$ as the expression obtained by replacing $\Omega$ by $-\Omega$ in the result of the integral in \eq{rt1}. 
  My interest lies in comparing $A(-\Omega)$ with $A(\Omega)$. If they are equal then the amplitudes for the particle to propagate with an energy $\Omega$ or $-\Omega$ are the same; but when they are unequal it indicates some interesting physics.
  
  To probe this issue, let us consider the explicit form of $G(\tau)$ in a $D$-dimensional flat spacetime given by (with $m^2$ treated as $m^2-i\epsilon$):
  \begin{equation}
 G(\tau) = i \frab{1}{4\pi i}^{D/2} \int_0^\infty \frac{ds}{s^{D/2}} \ \exp\left[-ism^2 - \frac{i}{4s}\, \sigma^2(\tau)\right]
   \label{rt2}
  \end{equation} 
  where $\sigma^2(\tau) \equiv \sigma^2 (x_1,x_2)$ is the squared line interval between the two events. The fact that $\sigma^2$ depends only on $\tau = \tau_1 - \tau_2$ again arises from the stationarity of the background and the existence of the Killing vector $\partial/\partial \tau$. From the structure of the integral in \eq{rt1} it is obvious that, if $G(\tau) = G(-\tau)$, then $A(\Omega) = A(-\Omega)$ so that nothing very interesting happens. This is, of course, trivially true if we take $\tau $ to be the standard inertial time coordinate $t$ so that $\sigma^2(t) = t^2 - |\bm{x}_1 - \bm{x}_2|^2$. This makes $\sigma^2$ and $G$  even functions of the time difference, leading to $A(\Omega ) = A(-\Omega)$. 
  
  Interestingly enough, the same result holds even when both events $x_1$ and $x_2$ are on the right Rindler wedge (R) with $\tau$ being the Rindler time coordinate. In R the Rindler coordinates $(\tau, \rho)$ can be defined\footnote{We will work with units such that the acceleration $g$ of the Rindler frame is unity. In the coordinate transformation from $(t,x,\bm{x}_\perp)$ to $(\tau, \rho, \bm{x}_\perp)$,  the transverse coordinates $\bm{x}_\perp$ go for a ride and I will not display them unless necessary.} in the usual manner as $t=\rho \sinh \tau$, \ $x= \rho \cosh \tau$.   The line interval $\sigma^2_{RR}$ for two events in the right wedge has the form 
  \begin{equation}
 \sigma^2_{RR}(\tau) = - L_1^2 + 2 \rho_1\rho_2 \, \cosh \tau
   \label{rt3}
  \end{equation} 
  where 
  $
 L_1^2 = (\Delta \mathbf{x}_\perp^2 + 2 \xi_1 + 2 \xi_2)
  $,
  with the $\xi$ coordinate defined through the relation  $x^2 - t^2  \equiv 2\xi$. (In R, $2\xi=\rho^2$). The $\sigma^2_{RR}(\tau)$ is clearly an even function of $\tau$ and hence we reach the following conclusion: \textit{When a particle propagates between any two events within the right Rindler wedge R, we have $A(\Omega) = A(-\Omega)$.}\footnote{The Unruh-Dewitt detector response \cite{detector}, for a uniformly accelerated trajectory in R, is computed by a Fourier transform similar to the one in \eq{rt1}, for events  with $\rho_2 = \rho_1$,\ $\Delta \bm{x}_\perp =0$, with the \textit{Wightman function} replacing the propagator. This, of course, leads to $A(-\Omega)\neq A(\Omega)$.  The difference arises due to the difference in the structure of Wightman function and the Feynman propagator. Algebraically, $[\sinh^2(\tau/2) -i\epsilon]$ --- which occurs in the propagator --- is an even function of $\tau$
  while $\sinh^2[(\tau/2) -i\epsilon]$ --- which occurs in the Wightman function --- is not.}
  
  Let us next consider what happens when we take one event to be in R and the second event to be in F where the Rindler-like coordinate system is introduced through $t=\rho \cosh \tau$ and $x=\rho \sinh \tau$. (If one uses the $\xi$ coordinate, then the relation $x^2 - t^2 = 2\xi$ allows the region F to be covered by the range $-\infty < \xi < 0$ and the region R to be covered by the range $0<\xi<\infty$.) The line interval $\sigma^2_{FR} (\tau)$ between an event $(\tau_F, \rho_F) $ in F and an event $(\tau_R, \rho_R) $ in R is  given by
  \begin{eqnarray}
 \sigma^2_{FR}(\tau)&\equiv& (t_F-t_R)^2-(x_F-x_R)^2-\Delta \mathbf{x}_\perp^2\\
 &=&(\rho_F\cosh\tau_F- \rho_R\sinh\tau_R)^2 - (\rho_F\sinh\tau_F- \rho_R\cosh\tau_R)^2-\Delta \mathbf{x}_\perp^2\\
 &=&\rho_F^2-\rho_R^2-2 \rho_F\rho_R\sinh(\tau_R- \tau_F)-\Delta \mathbf{x}_\perp^2
 \label{vpone}\\
 &\equiv&- L_2^2 - 2 \rho_F\rho_R \, \sinh \tau; \qquad\qquad \tau\equiv(\tau_R-\tau_F)
 \label{tp1}
 \end{eqnarray} 
 with  $L_2^2 \equiv (\Delta \mathbf{x}_\perp^2 + 2 \xi_R + 2| \xi_F|)$.
 I displayed this calculation in gory detail because there is a bit of algebraic sorcery involved in it. (This is the only non-trivial calculation in this paper!) The line interval $\sigma^2(\mathcal{P}_1,\mathcal{P}_2)$ between any two events in the spacetime, of course, is symmetric with respect to the interchange of events,
 $\sigma^2(\mathcal{P}_1,\mathcal{P}_2)=\sigma^2(\mathcal{P}_2,\mathcal{P}_1)$. In our case, the two events have the  coordinates 
 \begin{equation}
\mathcal{P}_1=\mathcal{P}_F=(t_F,x_F,\mathbf{x}_F^\perp)= (\rho_F\cosh\tau_F,\rho_F\sinh\tau_F, \mathbf{x}_F^\perp) 
\label{p1co}
\end{equation} 
 and 
 \begin{equation}
\mathcal{P}_2=\mathcal{P}_R=(t_R,x_R, \mathbf{x}_R^\perp)= (\rho_R\sinh\tau_R,\rho_R\cosh\tau_R,\mathbf{x}_R^\perp).
\label{p2co}
\end{equation}  
 The symmetry of the line interval is manifest in the inertial coordinates and we have $\sigma^2(t_F,\mathbf{x}_F;,t_R,\mathbf{x}_R)=\sigma^2(t_R,\mathbf{x}_R;,t_R,\mathbf{x}_F)$. But you \textit{cannot} display the same symmetry by interchanging the relevant Rindler coordinates!  From the \eq{vpone} we see that 
 \begin{equation}
 \sigma^2(\tau_F,\rho_F,\mathbf{x}_F^\perp;\tau_R,\rho_R, \mathbf{x}_R^\perp)\neq\sigma^2(\tau_R,\rho_R,\mathbf{x}_R^\perp;\tau_F,\rho_F, \mathbf{x}_F^\perp)
  \end{equation} 
 Of course, if you introduce arbitrary coordinate labels to events in spacetime,  there is no assurance that the interchange of coordinate labels will correspond to the interchange of events, when two different coordinate charts are involved. This is precisely what happens here: It is obvious from \eq{p1co} and \eq{p2co} that the interchange $(\tau_F,\rho_F)\Leftrightarrow(\tau_R,\rho_R)$, of coordinate labels we are using,  does \textit{not} lead to the interchange of the events $\mathcal{P}_1\Leftrightarrow \mathcal{P}_2$ because two different coordinate charts\footnote{How come $\sigma^2$ between the events in R and F only depends on the difference in `time' labels, especially since $\tau$ is not even a time variable in F? This has to do with the fact that one can indeed introduce,  a (Schwarzchild-like) coordinate system covering both R and F in which the 2-D metric takes the form $ds^2=(2\xi)d\tau^2-(2\xi)^{-1}d\xi^2$. We see that $\tau$ retains its Killing character both in R and F, though $\partial/\partial\tau$ is timelike only in R. It is the Killing character which ensures that $\sigma_{FR}^2$ only depends on the difference in the `time' labels.} are used in R and F.
 
We will now compute the Fourier transform in \eq{rt1} with  respect to $\tau\equiv(\tau_R-\tau_F)$. The sign convention in \eq{rt1} implies that $G$ picks up a contribution $A(\Omega)\exp -i\Omega(\tau_R-\tau_F)$ which will correspond to a positive energy with respect to $\tau_R$ when $\Omega>0$ (and negative energy when $\Omega<0)$. These are defined with respect to $\tau_R$ which is a valid time coordinate in R. (So I do not have to worry about the fact that $\tau_F$ has no clear meaning as a time coordinate in F; it is an ignorable constant which goes away when I do the integral over the range $-\infty<\tau<\infty$.)   
 The Fourier transform in \eq{rt1} requires us to compute the integral:
   \begin{equation}
 I = \int_{-\infty}^\infty d\tau\ e^{i\Omega \tau - \frac{i}{4s} \sigma_{FR}^2 (\tau) }
 = 2\, e^{\frac{iL^2}{4s} } \ e^{-\pi \Omega/2} K_{i\Omega}(2\alpha)
   \label{rt7}
  \end{equation}
  where $\alpha \equiv  (\rho_1\rho_2/2s)$. This was done using the standard integral representation for the McDonald function, leading to:
   \begin{equation}
  \int_0^\infty \frac{dq}{q} \, q^{i\omega} \, e^{i\alpha\left( q - \frac{1}{q}\right)} = 2\, e^{-\pi\omega/2}\ K_{i\omega} (2\alpha); \qquad (\alpha>0)
   \label{rt8}
  \end{equation}
 Substituting \eq{rt7} into \eq{rt1}, we find that the relevant amplitude is given by 
 \begin{equation}
 A(\Omega) = e^{-\pi\Omega/2} \, \int_0^\infty ds\ F(s) K_{i\Omega}(2\alpha)
   \label{rt9}
  \end{equation}
  where 
  \begin{equation}
  F(s) = 2i \frab{1}{4\pi is}^{D/2}\, e^{-im^2 s + \frac{iL_2^2}{4s}}
   \label{rt10}
  \end{equation}
  Since $K_{i\Omega} = K_{-i\Omega}$ is an even function of $\Omega$, it follows that 
  \begin{equation}
  A(-\Omega) = e^{\pi\Omega/2} \, \int_0^\infty ds\ F(s) K_{i\Omega}(2\alpha) = e^{\pi \Omega}\ A(\Omega)
   \label{rt11}
  \end{equation}
  leading to the familiar Boltzmann factor 
  \begin{equation}
  \frac{|A(\Omega)|^2}{|A(-\Omega)|^2} = e^{-2\pi\Omega}
   \label{rt12}
  \end{equation}
  corresponding to the Davis-Unruh\cite{davies-unruh} temperature $T=g/2\pi = 1/2\pi$ in our units.\footnote{The analysis leads to similar conclusions for other situations when the events are separated by a horizon, like for e.g., between region P and region L. I will concentrate on F and R.}  
  This result is equivalent to attributing a temperature $T=1/2\pi$ to the horizon when viewed from R.
  The propagation of a particle with energy $\Omega$ from a spatial location in F to a spatial location in R can be thought of as an emission of a particle by the horizon surface, since an observer confined to R cannot (classically) detect anything beyond the horizon. By the same token, the propagation of a particle with an energy $-\Omega$ can be thought of as the absorption of energy $\Omega$ by the horizon. Therefore, we have $P_e/P_a = |A(\Omega)|^2/|A(-\Omega)|^2$ where $P_e, P_a$ denote the probabilities for emission and absorption.
   On the other hand, if we think of the horizon as a hot surface, with fictitious two-level systems  in thermal equilibrium, then $P_e \propto N_{\rm up}$ and $P_a \propto N_{\rm down}$ where $ N_{\rm up}$ and $ N_{\rm down}$ are the population of the upper and lower levels separated by energy $\Omega$. Therefore our result in \eq{rt12}
implies that  $ N_{\rm up}/ N_{\rm down} = e^{-2\pi \Omega}$ showing that the level population of two-level system, on the horizon surface, satisfies the Boltzmann distribution corresponding to the temperature $T= 1/2\pi$. 
  This is a more concrete, physical, interpretation of the result in \eq{rt12}.
  
  I find it particularly gratifying that the propagator can distinguish so nicely between the propagation across the horizon from the propagation within one side of the horizon. Let me stress how this fact prevents you from interpreting (`understanding') the \eq{rt12} in a trivial manner: You might think, at first sight, that if I am Fourier transforming $G$ with respect to the \textit{Rindler time} $\tau$ (and define positive/negative energies through $\exp\mp i\Omega\tau$) then it is a foregone conclusion that I will get the thermal factor. \textit{This is simply not true.} Recall that, when I do the Fourier transform with respect to Rindler time etc. but for two events within the right wedge R, I do not get a thermal factor. So the usual suspect, viz., $\exp-i\Omega t$ being a superposition of $\exp\mp i\Omega \tau$), is \textit{not} responsible for this result. There are two other crucial ingredients which go into it. First, you need horizon crossing to break the symmetry between $G(\tau)$ and $G(-\tau)$; this is obtained, as I said, by the only non-trivial calculation in this paper, leading to \eq{tp1}. Second, it is crucial that the result in \eq{tp1} depends only on the difference $\tau\equiv(\tau_R -\tau_F)$. So  when I integrate over all $\tau$, I don't have to worry what $\tau_F$ means, since it is not a time coordinate in F. I can stay in R and interpret everything using $\tau_R$. Therefore, it is not just using the Rindler time coordinate which leads to the result. The structure of the propagator is more nontrivial than one would first imagine.
  
  As far as I know, this is the first piece of work which obtains the 
  thermality of the Rindler horizon directly from the \textit{propagation amplitude across the horizon} in 
  a clean, direct manner, without using Rindler modes, Rindler quantization, 
  Rindler vacuum etc. 
  To do this, I \textit{have to} use  the Feynman propagator which describes the propagation amplitude;  other
  two-point functions can describe \textit{vacuum correlations} but they do \textit{not} describe \textit{propagation amplitude}.
  Previous attempts have obtained  thermality, either by extracting the spectrum of vacuum fluctuations
  (as in detector response) or by studying the  entanglement and correlations between R and F wedges (see \cite{higuchi} as well as \cite{mastas}).
To make the conceptual difference between these attempts clearer, let me emphasize  the  physical distinction between vacuum correlations (represented by two-point-functions like Wightman function $G^+(x_1,x_2)$), and propagation (represented by the Feynman propagator  $G^(x_1,x_2)$).
  
What is crucial for describing \textit{relativistic} propagation is the following fact: Feynman propagator evolves positive frequencies forward in time and negative frequencies backwards in time. (For a text book elaboration, see e.g., Sec 1.5.1 of  Ref.\cite{tpqft}.) 
The spatial Fourier transform $G(t,\bm{k})$ [of the Feynman propagator $G(t,\bm{x})$]  has the factor $\exp(-i\omega_{\bm{k}} |t|)$ where $\omega_{\bm{k}}=+\sqrt{\bm{k}^2+m^2}$. (See e.g., eq (1.85) of Ref. \cite{tpqft}). The modulus sign in $|t|$ is crucial for the interpretation which, in turn, is equivalent to time-ordering of $\phi(x)\phi(y)$ in the VEV.  On the other hand, the spatial Fourier transform of the Wightman function $G^+$ has the factor $\exp(-i\omega_{\bm{k}}t)$   
(without modulus on $t$) and hence it only has forward-in-time evolution; similarly $G^-$ only has backward-in-time evolution.\footnote{This difference is very apparent in  the case of a \textit{complex} scalar field but it exists, of course,  for the \textit{real} scalar field as well; for a complex field, written as $\phi(x)\equiv A(x)+B^\dagger(x)$ --- where $A(x)$ and $B(x)$ are made of positive frequency modes --- the  $G(x_1,x_2)= \bk{M}{A(x_1)A^\dagger(x_2)}{M}$ if $x_1^0>x_2^0$ while it is $G(x_1,x_2)= \bk{M}{B(x_1)B^\dagger(x_2)}{M}$ if $x_1^0<x_2^0$; here, $\ket{M}$ is the inertial vacuum state. On the other hand, $G^+(x_1,x_2)$ is always $\bk{M}{A(x_1)A^\dagger(x_2)}{M}$ and misses the antiparticle (`backward in time') propagation, contained in $B(x)$. That piece of information is contained in the complementary function $G^-(x_1,x_2)$
  which will always be $\bk{M}{B(x_1)B^\dagger(x_2)}{M}$ thereby missing the particle (`forward in time') propagation; the Feynman propagator has both pieces of information.}
  Therefore, in QFT, to study the \textit{propagation} of a particle between events (especially across the horizon) we \textit{must} use $G(x_1,x_2)$; the Wightman function  $G^+(x_1,x_2)$ is insufficient and inappropriate. This is also obvious from the following two facts  stated right at the beginning of the paper: 
  
 (a) The path integral for a relativistic particle in Eq. (1) sums over paths which go both backwards and forwards in time and leads naturally to $G(x_1,x_2)$ (and not to $G^+(x_1,x_2)$). Similarly, the path integral average of $\phi(x_1)\phi^\dagger(x_2)$ will also lead to $G(x_1,x_2)$.
  
  (b) The emission and absorption of particles by a source $J(x)$ in Eq. (2) are described using $G(x_1,x_2)$ (and not $G^+(x_1,x_2)$). This is linked to the crucial fact that, in QFT, any source which emits particles \textit{must} also absorb them, which forms a cornerstone of Schwinger's source theory. In our case, the emission and absorption of particles by the horizon involves `backward' propagation from F to R and hence \textit{has to be} discussed in terms of $G(x_1,x_2)$. 
  
  It is certainly  possible to obtain the Rindler temperature using $G^+(x_1,x_2)$ either (i) in the context of response of particle detectors or (ii) in terms of entanglement and correlations between R and F (see [4] as well as [5]).
  In the approach (i), horizon plays no role; detector will click in several trajectories which do not asymptote to a horizon because it merely records the spectrum of vacuum fluctuations encoded in $G^+(x_1,x_2)$. In the approach (ii), adopted in [4,5]  also no \textit{propagation} across the horizon is used or computed anywhere; in fact,  $G^+(x_1,x_2)$ is incapable of describing propagation. So, while this approach is interesting and provides a different, complementary, perspective of the horizon thermality, it is \textit{distinctly} different from the analysis presented here.
  
  In obtaining  this result, I worked entirely in the Lorentzian sector with a well-defined causal structure and the horizons at $x^2-t^2=0$. I have also emphasized the key role played by the horizon in obtaining  this result. You may wonder what happens to this analysis if it is done with the inertial  propagator in the \textit{Euclidean} sector. In the conventional approach, the right wedge (with $t=\rho\sinh\tau, x=\rho\cosh\tau$) itself will fill the \textit{entire} Euclidean plane $(t_E,x_E)$ if we take $it=t_E, i\tau=\tau_E$ leading to $t_E=\rho\sin\tau_E, x=\rho\cos\tau_E$. The horizons ($x^2-t^2=0$) map to the origin ($x^2+t_E^2=0$) and the F,P,L wedges seem to disappear! At first sight, it is not clear how to recover the information contained in the F,P,L wedges if we start with the Euclidean, inertial, propagator. However, it can be done but one needs to use four different types of analytic continuations to proceed from the Euclidean plane to the four Lorentzian sectors (R, F, L, P). I have described this briefly in Appendix A for the sake of completeness.
  
  \section{The horizon thermality hiding in the inertial propagator}
  
  Given these facts, let me probe the structure of the inertial propagator a little more closely.  While obtaining the above result I did not compute the final integral in \eq{rt9} because it was unnecessary. However, this can be done both for events in R and for two events separated by a horizon. The relevant integrals are simpler to exhibit if we first get rid of the transverse coordinates, by Fourier transforming both sides of \eq{rt1} 
  with respect to the transverse coordinate difference  ($\bm{x}^\perp_1-\bm{x}^\perp_2$), thereby introducing the conjugate variable $\bm{k_\perp}$. (As usual, I will simply write $G^{(RR)}(\tau)$ for $G(\tau; \rho,\rho'; \mathbf{k}_\perp)$ when both events are in R etc.)
  It can be shown  that, when both events are located in R, the relevant Fourier transform in \eq{rt1} is given by
  \begin{equation}
 A_{RR}(\Omega) = \int_{-\infty}^\infty d\tau\ G^{(RR)}(\tau) \, e^{i\Omega \tau} = \frac{i}{\pi} \, K_{i\Omega} (\mu \rho_2) \, K_{i\Omega}(-\mu \rho_1);\qquad \tau =(\tau_1 -\tau_2)
   \label{rt13}
  \end{equation}
 with the ordering,  $\rho_1<\rho_2$.  But if the events are in F and R the corresponding Fourier transform is
  \begin{equation}
 A_{FR}(\Omega) = \int_{-\infty}^\infty d\tau\ G^{(FR)}(\tau) \, e^{i\Omega \tau} = \frac{1}{2} \, H_{i\Omega}^{(2)} (\mu \rho_F) \, K_{i\Omega}(\mu \rho_R);\qquad \tau =(\tau_R -\tau_F)
   \label{rt14}
  \end{equation}
  where $\mu^2 \equiv k_\perp^2 + m^2$.
  (I have sketched the derivation in Appendix A. The result is also closely related to the form of the Minkowski-Bessel modes \cite{gerlach} in R and F.) The occurrence of the Hankel function $H_{i\Omega}^{(2)}$ in \eq{rt14} in contrast to the McDonald function in \eq{rt13} makes all the difference because --- while McDonald function is even in its index --- Hankel function has the property $H_{i\nu}^{(2)} = e^{-\pi \nu}  H_{-i\nu}^{(2)} $. This immediately gives 
  \begin{equation}
  \left[\frac{A(\Omega)}{A(-\Omega)}\right]_{FR} =\frac{H_{i\Omega}^{(2)}}{H_{-i\Omega}^{(2)}} = e^{-\pi\Omega}
   \label{rt15}
  \end{equation}
  which is the same as \eq{rt11}. On the other hand, because $K_{i\Omega}= K_{-i\Omega}$ we trivially get $A_{RR}(\Omega) = A_{RR}(-\Omega)$.
  So the explicit computation verifies the previous result but --- as I will argue later --- the original approach offers greater generality. 
  
  This is not the only manner in which the inertial propagator hides the thermal nature of the Rindler horizon. I will give one more example which actually takes you to the Rindler quantization --- something I have judiciously avoided so far --- from the structure of inertial propagator. To do this, let us start with the \textit{Euclidean} version of the inertial propagator for two events in R:
  \begin{equation}
 G_{\rm Eu}^{inertial} (\bm{k}_\perp ; \, \rho_1, \rho_2, \theta-\theta') = \frac{1}{2\pi^2} \int_{-\infty}^\infty d\nu\, e^{\pi \nu} \, K_{i\nu}(\mu \rho_2) \, K_{i\nu}(\mu\rho_1)\ e^{-\nu|\theta-\theta'|}
   \label{rt22new}
  \end{equation}
  As before, I have already Fourier transformed with respect to the transverse coordinate difference  ($\bm{x}^\perp_1-\bm{x}^\perp_2$) thereby introducing the conjugate variable $\bm{k_\perp}$. Further $\mu^2 = k_\perp^2 + m^2$.
 (This expression, with a $|\theta-\theta'|$ is well-known in literature and is very easy to derive. In Appendix A, I have given the derivation as well as its relation with the form in \eq{rt13}, which is based on another variant with  $(\theta-\theta')$; this one is  a bit nontrivial to derive.). Using just a series of Bessel function identities and \textit{no physics input}, this result can be re-expressed in the following form:
 \begin{equation}
  G_{\rm Eu}^{inertial} (\theta-\theta')=\sum_{n=-\infty}^{\infty} G_{\rm Eu}^{Rindler}[\theta-\theta'+2\pi n]
  \label{more1}
 \end{equation} 
 where the function $G_{\rm Eu}^{Rindler}$ is given by:
 \begin{equation}
G_{\rm Eu}^{Rindler} \equiv \frac{1}{\pi^2} \int_0^\infty d\omega\ (\sinh \pi \omega) \, K_{i\omega} (\mu \rho)\, K_{i\omega}(\mu\rho')\, e^{-\omega|\theta -\theta'|}
\label{more2}
\end{equation} 
This results tells us two things: (a) First, the Euclidean version of our standard inertial propagator can be expressed as an infinite, periodic sum in the (Euclideanised) Rindler time. The fact that inertial propagator is periodic in (Euclideanised) Rindler time is a trivial result; you only need to note that the $\sigma_{RR}^2$ in  \eq{rt3} is periodic in $i\tau$. But \eq{more1} and \eq{more2} give us a lot more information. They explicitly express $G_{\rm Eu}^{inertial}$ this an infinite periodic sum of \textit{another specific function}  $G_{\rm Eu}^{Rindler}$. (b) From the product structure of $G_{\rm Eu}^{Rindler}$, we learn that, when analytically continued back to Lorentzian sector,    
it can be thought of as a propagator built from another set of mode functions:  
 \begin{equation}
  \phi_\nu (\tau,\rho) = \frac{1}{\pi} \left( \sinh \pi \nu\right)^{1/2} \, K_{i\nu}(\mu \rho)e^{-i\nu \tau}
   \label{rt17}
  \end{equation}
 in the standard fashion with time ordering with respect to $\tau$. This allows us to discover the Rindler mode functions, Rindler vacuum and the Rindler propagator, just from analyzing the inertial propagator and rewriting it as in \eq{more1} and \eq{more2}. (Of course, the modes in \eq{rt17} satisfy the Klein-Gordon equation and are properly normalized.) So just staring at the inertial propagator, you can discover the Rindler modes and the Rindler vacuum.
 
 There is another, closely related, feature. To bring this out, I
   will introduce a \textit{reflected} wave function $\phi_\nu^{(r)}$ by the definition 
  \begin{equation}
  \phi_\nu^{(r)} (\rho,\tau) = \phi_\nu (-\rho,\tau-i\pi) = \phi_\nu(\rho^r, \tau^r)
   \label{rt19}
  \end{equation}
  The adjective ``reflected''  is justified by the facts that: (i) The coordinates $\rho$ and $-\rho$ are obtained by a reflection through the origin and (ii) the replacement of $\tau$ by $\tau-i\pi$  in the Rindler coordinate transformation takes you from R to L. (If you replace $\rho$ by $-\rho$ and \textit{also replace} $\tau$ by $\tau-i\pi$ in the coordinate relations ($x=\rho\cosh\tau, t=\rho\sinh\tau$), you will get back  to the same event in $R$. But  $\phi_\nu^{(r)} (\rho,\tau)\neq\phi_\nu (\rho,\tau)$, making the reflected wave function different from the original one.) 
  It turns out that  propagator for two events \textit{within the right wedge} can be expressed in a very suggestive form as:\footnote{The proofs for all these claims, like e.g., \eq{rt22new}, \eq{more1}, \eq{more2} are sketched in Appendix A}  
  \begin{equation}
 G^{(RR)} = \int_0^\infty d\nu\, \left[ (n_\nu + 1) \, \phi_\nu\, \phi_\nu^{(r)} + n_\nu \phi_\nu^*\, \phi_\nu^{(r)*}\right]
   \label{rt20}
  \end{equation}
  where $n_\nu$ is the thermal population: 
  \begin{equation}
  n_\nu = \frac{1}{e^{2\pi \nu} - 1}
   \label{rt21}
  \end{equation}
  Obviously, the second term in \eq{rt20} suggests  an absorption process weighted by $n_\nu$  while the first term could represent emission with the factor $n_\nu +1$ coming from a combination of stimulated emission and spontaneous emission.  If we think of $\phi_\nu$  and $\phi_\nu^r$  as the wave functions for a fictitious particle, then this structure again encodes the usual thermality.\footnote{in the usual approach, the  Bogoluibov transformation between inertial and Rindler modes involves $|\beta|^2\sim n_\nu,
  |\alpha|^2\sim (1+n_\nu)$ and one can transform $G(x_1,x_2)=\bk{0}{T[\phi(x_1)\phi(x_2)]}{0}$, expressed in inertial modes to one involving Rindler modes. This is a way of connecting up \eq{rt20} to something more familiar.
  The factors multiplying $(1+n)$ and $n$ can be related to the Bremsstrahlung by an accelerating source. In fact, both terms will correspond to emission when viewed in the inertial frame.}
  
  Since the Rindler frame is just a coordinate transformation of the inertial frame and the propagator $G(x_1,x_2)$ transforms as a bi-scalar under coordinate transformation, we can trivially  represent it  in Rindler coordinates. Further because $G(x_1,x_2)$ encodes all the physics contained in a free  field we should be able to discover the thermality just by staring at $G(x_1,x_2)$. In other words, it should not be necessary for me to quantize the field in Rindler coordinates, identify positive frequency modes, construct Rindler vacuum and particles etc etc. Everything should flow out of $G(x_1,x_2)$ expressed in Rindler coordinates including the alternative, Rindler, quantization. This is what I have achieved in the above discussion. 
  
  \section{Discussion}
  
  \subsection{Comparison with other approaches}
  
  There are  three other main approaches which follows similar philosophy --- viz., to obtain the Davies-Unruh temperature without using explicit Rindler quantization --- as far as thermality of the horizon is concerned. (None of them, however, takes you beyond that, to the results I have obtained in Section 2.) The first one is through the response of Unruh-Dewitt detector in which one merely calculates a Fourier transform of the Wightman function. The second is the path integral approach used in \cite{hh}. Finally, the horizon tunneling approach (see, for example, \cite{tunneling}) has some superficial similarity with the ideas presented above.

  The approach in Section 1 of this paper is quite different from all the three approaches mentioned above.  To begin with, it makes use of the Feynman propagator, the central quantity in QFT, and obtains the thermality from it. I stress that the Feynman propagator has a hidden structure which ensures that the notion of thermality arises when one events are separated by a horizon but not otherwise.  So the `horizon crossing' plays a crucial but hidden role. This is not the case with the calculation of the detector response. It is not obvious (in a calculation confined within R) what exactly is the role played by the horizon, if any. In fact a detector in any non-trivial trajectory will click --- albeit in a complicated and time dependent manner --- even if there is no horizon. So the superficial similarity --- of evaluating a Fourier transform of a two-point function --- should not mislead you in this matter. 
  
  The path integral approach in \cite{hh}, again, has a superficial similarity with what I have done here. However, there are some significant differences. First, the derivation in \cite{hh} suggests that the probability, for the absorption of  a particle by a region beyond the horizon, is  related by a thermal factor to the probability for the emission from that region. This is very different from the interpretation I am trying to advocate. I just look at the propagation amplitude $A(\Omega)$ in energy domain and ask how $A(\Omega)$ and $A(-\Omega)$ are related, for propagation between the same pair of events. I have to again stress that the non-trivial structure  of the Feynman propagator ensures that when the events are separated by a horizon a thermal relationship arises. Second, the analysis in \cite{hh} crucially uses the white hole region (P) to arrive at the conclusion.  My approach just uses F and R  and hence is conceptually clearer. 
  
 Finally, my approach is quite distinct from the standard lore of deriving thermality from horizon tunneling. First, the tunneling approach --- like the path integral approach ---  tries to relate the amplitude for absorption by F to the  emission from F and claims that these two are different because of the pole structure in complex plane. I did not have to resort procedures like analytic continuation in the main derivation. Further, it is not very clear how structure of quantum field theory --- encoded in the propagator --- is incorporated in the tunneling  approach. In contrast, it is very clear in what I have done.  
  
  \subsection{Generalizations}
  
  The approach, and the result, have obvious generalizations to more complicated situations and I concentrated on the Rindler thermality only for keeping things simple. To begin with, the result can be extended to de Sitter spacetime in a straight forward manner because the dependence of the propagator on the geodesic distance (see, for e.g., \cite{kkvp}) allows the same derivation to go through.\footnote{In the case curved spacetimes with horizons, like in Schwarzchild, Reissner-Nordstrom etc. we get the same result by explicit computation in $D=2$. In $D>2$, we do not have closed expressions for $G(x,x')$, but one can compute it close to the horizon. This is because, close to the horizon, you again get a 2D CFT and one can compute approximate form of the modes --- and through them --- the propagator $G(x,x')$. This will lead to the same result.}
 More generally, one can use this approach to attribute thermality to any \textit{local Rindler horizon} along the following lines. 
 
 In an arbitrary spacetime, pick an event $\mathcal{P}$ and introduce the Riemann normal coordinates around $\mathcal{P}$. These coordinates will be valid in a region, $\mathcal{V}$, of size $L$ where the typical background curvature is of the order of $L^{-2}$. Introduce now a local Rindler coordinate system by boosting with an acceleration $g$ with respect to the local inertial frame, defined in $\mathcal{V}$. If we now concentrate on events  $(x_1, x_2)$ within $\mathcal{V}$, then the standard Schwinger-DeWitt expansion of the propagator tells us that the form in \eq{rt2} will be (approximately) valid. The Fourier integral in \eq{rt1} can be defined formally, though the range of $\tau$ outside the domain $\mathcal{V}$ is not meaningful. To circumvent this, we have to arrange matters such that most of the contribution to the integral in \eq{rt1} comes from the range  $\tau \lesssim L$. This, in turn, requires us to concentrate on the high frequencies with $\Omega \gg L^{-1}$. In this high frequency limit everything will go through as before and one will obtain the local Rindler temperature to be $T = g/2\pi$. For consistency, we also need to ensure that $gL \gg 1$ which, of course, can be done around any event with finite $L$. (In fact, this approach suggests a procedure for obtaining the curvature corrections to the temperature systematically, using the Schwinger-DeWitt expansion.) I stress that --- in this very general context  of a bifurcate Killing horizon,   introduced into a local inertial frame --- my approach gets you whatever you could reasonably expect. After all, in a curved spacetime, one can expect  thermality (with approximately constant temperature)  only when the modes do not probe the curvature scale; this is what is achieved by concentrating on the Feynman propagator at two events which are localized within $\mathcal{V}$. 
 
 \subsection{Future directions}
 
 There are three  avenues of further work which seem interesting. First is to probe the uniqueness of the result in \eq{more1} and \eq{more2}. I have shown that, starting from just the Euclidean version of the inertial propagator and the coordinate transformation in the right wedge, one can obtain \eq{more1} and \eq{more2}. This is just Bessel function gymnastics with \textit{no physics input.} But the resulting structure in \eq{more2} --- involving product of mode functions and time ordering with respect to $\tau$, when analytically continued back into Lorentzian sector --- immediately suggests an alternative set of mode functions (with positive/negative frequency decomposition with respect to $\tau$), corresponding Rindler vacuum and the Rindler propagator. Then \eq{more1} tells us that inertial vacuum will appear as a thermal state in the new representation. Only thing missing is a proof that the form of the infinite periodic sum in \eq{more1} and \eq{more2} is unique. I think this is true but might require some analyticity assumptions. 
 
 Second, one might like to probe 
 the details of emission/absorption by localized sources (e.g, on two sides of a horizon) using the expression in \eq{rts1} and connecting up with the structure in \eq{rt20}.  This will throw more light on how such processes appear in inertial coordinates versus Rindler coordinates. In fact, I expect both processes to appear as emission in the inertial frame.
 
Third, it will be interesting to see whether the path integral in \eq{rt0} can be computed from first principles in the Rindler coordinates. It can be done (even with a non-quadratic action) in inertial coordinates by a lattice regularization \cite{pilattice}. But it is not clear how to introduce a suitable lattice, either in polar coordinates in the Euclidean sector, or in the Rindler frame in the Lorentzian sector. These and related issues are under investigation.
 
\section*{Acknowledgement}

I thank Karthik Rajeev for discussions and comments on the draft. My research  is partially supported by the J.C.Bose Fellowship of Department of Science and Technology, Government of India.

  \section*{Appendix A: The unreasonable effectiveness of the Euclidean continuation}
  
  I will briefly outline the steps involved in obtaining \eq{rt13}, \eq{rt14}, \eq{more1}, \eq{more2}, \eq{rt20} and some related results, postponing their detailed discussion to another publication. \textit{I will now use mostly positive signature so that the analytic continuation of the time coordinate leads to a positive definite metric.}
  
  One can obtain  \eq{rt13} and  \eq{rt14} by doing the remaining integral in \eq{rt9} (and the analogous one for RR case) but this requires fairly complicated manipulation of known integrals over Bessel functions. But, since I also want to describe how to do the analytic continuation from the Euclidean sector to get all the four wedges (R, F, L, P), I will follow an alternative route. I will start from the \textit{Euclidean} propagator and obtain all the relevant results we need by careful analytic continuation. 
  
  The Euclidean (inertial) propagator  can be expressed in polar coordinates (with $x=\rho \cos \theta, \ t_E = \rho \sin\theta$)  in the following form
  \begin{equation}
 G_{\rm Eu} (\bm{k}_\perp ; \, \rho_1, \rho_2, \theta) = \frac{1}{2\pi^2} \int_{-\infty}^\infty d\nu\, e^{\pi \nu} \, K_{i\nu}(\mu \rho_2) \, K_{i\nu}(\mu\rho_1)\ e^{-\nu|\theta|}
   \label{rt22}
  \end{equation}
  In obtaining this propagator, I have already Fourier transformed with respect to the transverse coordinate difference  ($\bm{x}^\perp_1-\bm{x}^\perp_2$) thereby introducing the conjugate variable $\bm{k_\perp}$. Further $\mu^2 = k_\perp^2 + m^2$. 
  This result is well known in literature and is trivial to obtain. One begins by noting that if you Fourier transform the transverse coordinates in the Euclidean version of the propagator in \eq{rt2} you just get the reduced (two-dimensional) propagator, viz. $K_0 (\mu\ell)/2\pi$ where 
  $\ell = |\bm{\rho}_1 - \bm{\rho}_2|$.
  One can then use a standard identity
  \begin{equation}
   \frac{1}{2\pi} K_0(\mu \ell) = \frac{1}{\pi^2}\int_0^\infty d\nu\ K_{i\nu} (\mu \rho_1)\ K_{i\nu} (\mu \rho_2)\, \cosh[\nu(\pi - |\theta|)]
   \label{stdid}
  \end{equation}
  to express it as an integral over the range $0<\nu<\infty$.
  Extending the integration range  to  $(-\infty < \nu < \infty)$ we obtain \eq{rt22}. 
  
  To proceed from \eq{stdid} (which has $|\theta_1-
  \theta_2|$) to \eq{rt13} or \eq{rt14} (which have $(\theta_1-
  \theta_2)$), one needs to do the analytic continuation of the variables in a specific way. 
  Let me start with the approach to obtain \eq{rt13}.
  Usually one does the analytic continuation by $\theta_1\to i\tau_1, \theta_2\to i\tau_2$ and interpret 
  $|\theta_1-\theta_2|$ as $i|\tau_1-\tau_2|$, transferring the ordering to $\tau$ coordinate. This, of course, will give the correct Lorentzian propagator but with an $\exp (-i\nu|\tau_1-\tau_2|)$ factor. To get $(\tau_1-\tau_2)$ without the modulus, we need to employ the following\footnote{It is straightforward to verify that the coordinates transforms correctly from the Euclidean Rindler to Lorentzian Rindler under this transformation. To get the correct $i\epsilon$ prescription in the Lorentzian sector, it is important to interpret $(-\rho_<)$ as the limit of  $\rho_<\exp[i(\pi-\epsilon)]$. This aspect has been noticed  previously, in a different context, in Ref. \cite{candelas}. }  analytic continuation: $(\rho_>,\theta)\to(\rho_>,i\tau)$ and $(\rho_<,\theta')\to(-\rho_<,\pi+i\tau)$ with the ordering $\rho_>\ >\ \rho_<$. For complex numbers, we will interpret the relative ordering in $|z-z'|$ based on the real parts. This leads to the nice result that we now 
  end up replacing
  \begin{equation}
 e^{\pi \nu - \nu |\theta - \theta'|} \Rightarrow e^{- i\nu (\tau - \tau')}
   \label{rt23}
  \end{equation}
  Substituting this into \eq{rt22}, one immediately obtains
  \begin{equation}.
  G_{Min} = \frac{i}{\pi} \int_{-\infty} ^\infty \frac{d\nu}{2\pi} \, K_{i\nu}(\mu\rho_>)\, K_{i\nu} ( -\mu\rho_<) \, e^{- i \nu (\tau - \tau')}
   \label{rt24}
  \end{equation}
 from which \eq{rt13} follows. This is a simple way to get the result.
 
 But if you don't like simplicity and feel this is  a bit too slick, let me show you how to get this result from published tables of integrals. You again begin by recalling that, when you Fourier transform with respect to transverse coordinates in the \textit{Lorentzian} propagator, you  get the two-dimensional result $G_{Min}=iK_0(\mu\ell)/2\pi$ with 
 $
 \ell^2=\rho_{<}^2+\rho_{>}^2-2\rho_{<}\rho_{>}\cosh(\tau_2-\tau_1)
 $
 where we have ordered the $\rho$-s as $\rho_>\ >\ \rho_<$ for future convenience. (The  $\tau$ ordering is irrelevant; note that, in \eq{rt24}, interchanging $\tau$ and $\tau'$ corresponds to reversing the sign of $\nu$ which makes no difference because $K_{i\nu}$ is an even function of $\nu$.) Next, you
 look up the integral 6.792 (2) of \cite{gr} which gives, as a special case, the result:
 \begin{align}
	\int_{-\infty}^{\infty}\frac{d\omega}{\pi}e^{-i\omega \tau}K_{i\omega}(a)K_{i\omega}(b)= K_{0}(\sqrt{a^2+b^2+2ab\cosh\tau}); \quad(|\arg[a]|+|\arg[b]|+|\textrm{Im}[\tau]|<\pi)
	\label{spe1}
\end{align}
The left hand side almost looks like what we want but in the right hand side, the argument of $K_0$ has a term with $(+\cosh\tau)$ while our $\ell^2$ has $(-\cosh\tau)$. We need to take care of this and also ensure that $\sigma^2$ comes up as the limit of $\sigma^2+i\epsilon$ in the Lorentzian sector (i.e, Im$(\sigma^2)>0$).  To this end, make
the following identification in \eq{spe1}:
\begin{align}
	a=\mu\rho_{<}e^{i(\pi-\epsilon)};\qquad b=\mu\rho_{>}
\end{align}
with real $\tau$. Then  we have $|\arg[a]|+|\arg[b]|+|\textrm{Im}[\tau]|=\pi-\epsilon<\pi$ taking care of the condition in \eq{spe1}. 
Further, you can verify that the ordering $\rho_>\ >\ \rho_<$ also ensures that  Im$(\ell^2)>0$ leading to the correct $i\epsilon$ prescription in the Lorentzian sector. (The sign of imaginary part is decided by the sign of $(\rho_{>}\cosh(\tau)-\rho_{<})$ which remains positive due to our ordering of $\rho$-s.)
We thus get our advertised result:
\begin{align}
	\frac{i}{2\pi^2}\int_{-\infty}^{\infty}d\omega\ e^{-i\omega \tau}K_{i\omega}\left(-\mu\rho_{<}\right)K_{i\omega}(\rho_{>})= \frac{i}{2\pi}K_{0}\left(\mu\ell\right)=G_{Min}
\end{align}
which is the same as \eq{rt24}. I prefer the simpler derivation, though.
 
 To obtain the structure in \eq{rt14} we need to know how to proceed from the Euclidean sector to the wedge F. This is nontrivial because,  in the usual procedure of analytic continuation ($\theta\to i\tau$) you go from $(\rho \sin\theta, \rho \cos\theta)$ to 
 $(i\rho \sinh\tau, \rho \cosh\tau)$ which only covers the right wedge! But one can actually get all the four wedges from the Euclidean sector by using the following four sets of analytic continuations. (This is discussed in greater detail in \cite{rk-tp}.):
 \begin{align}
  &R: \ \rho \to \rho,\ \theta\to i\tau\,; && x=\rho \cosh \tau, \ t=\rho \sinh\tau\label{start}\\
  &F: \ \rho \to i\rho,\ \theta\to i\tau + \frac{\pi}{2}\,; && x=\rho \sinh \tau, \ t=\rho \cosh\tau\\
  &L: \ \rho \to \rho, \ \theta\to i\tau -\pi\,; && x=-\rho \cosh \tau, \ t=-\rho \sinh\tau\\
  &P: \ \rho \to i\rho, \theta =  i\tau - \frac{\pi}{2}\,; && x=-\rho \sinh \tau, \ t=-\rho \cosh\tau\label{end}
 \end{align} 
 Now using in R,  $(\rho,\theta)\to (\rho, i\tau)$ and using in F,
 $(\rho,\theta)\to (i\rho, i\tau+\pi/2)$ along with the identity
 \begin{equation}
  K_{i\nu} (iz) = -\frac{i\pi}{2}\ e^{-\pi \nu/2} \ H_{-i\nu}^{(2)} (z) =-\frac{i\pi}{2}\ e^{\pi \nu/2} H_{i\nu}^{(2)} (z)
 \end{equation} 
 one obtains a result similar to \eq{rt24} with a Hankel function replacing one McDonald function. This gives you \eq{rt14}. 
 
 In fact the analytic continuations in \eq{start} to \eq{end} allow us to obtain the propagator for any pair of points located in any two wedges directly --- and rather easily --- from the Euclidean propagator. You get a $K_{i\nu}K_{i\nu}$ structure in RR, LL, RL and LR. (The notation AB corresponds to first event being in wedge A and second in wedge B.) In FF, PP, FP and PF the McDonald functions are replaced by the Hankel functions. In PR, FL, RF, LP, RP and LF you get a product of a Hankel and McDonald function. 
 The interchange of F with P or R with L reverses the sign of $\nu$; so does the interchange of the two events. The similarity in structure with Minkowski-Bessel modes \cite{gerlach} is obvious. (These results agree with the ones in \cite{boulware}, obtained by more complicated procedure, except for some inadvertent typos in \cite{boulware}). We will discuss this procedure and results in detail in another publication \cite{rk-tp}.
 
 You can now obtain \eq{rt20}, working in the Lorentzian sector, by some further  straightforward manipulations.
 One starts with \eq{rt24} and converts it to an integral  the range $(0<\nu<\infty )$. Then using the results 
 \begin{equation}
 n_\nu = \frac{e^{-\pi \nu}}{2\, \sinh \pi\nu}\, ; \qquad 1+n_\nu = \frac{e^{\pi \nu}}{2\, \sinh \pi\nu}
   \label{rt25}
  \end{equation}
  we can rewrite the propagator as
  \begin{eqnarray}
  G^{(RR)} &=& \frac{i}{\pi^2} \int_0^\infty d\nu \,  K_{i\nu}(\mu\rho_>)\, K_{i\nu} ( -\mu\rho_<)
  \sinh \pi\nu \left[ e^{-\pi \nu} \, (n_\nu+1)\ e^{-i\nu \tau} + n_\nu\, e^{\pi \nu} \, e^{i\nu \tau} \right]\nonumber\\
  &=& \frac{i}{\pi^2} \int_0^\infty d\nu \,  K_{i\nu}(\mu\rho_>)\, K_{i\nu} ( -\mu\rho_<)
  \sinh \pi\nu \left[ (n_\nu+1) e^{-i\nu(\tau-i\pi)} + n_\nu e^{i\nu(\tau-i\pi)} \right]
   \label{rt26}
  \end{eqnarray}
   The pre-factors (outside the square bracket) lead to the product of wave functions in \eq{rt20} and the shift $(\tau-i\pi)$ leads to the reflected coordinate.

 However, the thermal factor in \eq{rt20} finds a more natural home in the Euclidean sector. Let me show you how this comes about --- using again  a set of identities related to Bessel functions ---  when we work in the Euclidean sector. First, the Euclidean propagator $K_0 (\mu\ell)/2\pi$ (obtained after transverse coordinates are removed by a Fourier transform) satisfies a Bessel function addition theorem (see page 351 (8) of \cite{watson})
 given by:
 \begin{equation}
G_E= \frac{1}{2\pi} K_0 (\mu \ell) = \frac{1}{2\pi} \sum_{m=-\infty}^\infty K_m (\mu \rho_{_>})\, I_m (\mu\rho_{_<}) \, \cos m (\theta - \theta')
 \end{equation} 
 The $K_mI_m$ part of the above result can be rewritten  in terms of another identity 
 you can look up (see 6.794(10) of \cite{gr}):
 \begin{equation}
 \frac{2}{\pi^2} \int_0^\infty d\omega \ \omega \, \sinh \pi \omega \, \frac{K_{i\omega}(\mu \rho) \, K_{i\omega}(\mu \rho')}{\omega^2+m^2} 
 = K_m(\mu \rho_{_>})\, I_m(\mu \rho_{_<})
 \label{tpeq2}
 \end{equation}
reaching: 
 \begin{equation}
G_E = \frac{1}{\pi^3} \sum_{m=-\infty}^\infty  \int_0^\infty d\omega \ \omega \, \sinh \pi \omega \, \frac{K_{i\omega}(\mu \rho) \, K_{i\omega}(\mu \rho')}{\omega^2+m^2}\ \cos m (\theta - \theta')
 \label{tpeq3}
 \end{equation}
The sum in the above expression can again be looked up (see 1.445 (2) of \cite{gr}); it is precisely the thermal factor in \eq{rt20} written in Euclidean sector:
\begin{eqnarray}
\mathcal{T}_\omega(\theta-\theta')&\equiv& \sum_{m=-\infty}^\infty\frac{1}{\pi}\frac{\omega}{\omega^2+m^2}\ \cos m (\theta - \theta')
=\frac{\cosh\omega(\pi-|\theta - \theta'|)}{\sinh\pi\omega}\nonumber\\
&=&(n_\omega+1)e^{-\omega|\theta - \theta'|}+n_\omega e^{\omega|\theta - \theta'|}
\label{tpeq3new}
 \end{eqnarray}
This will lead to the Euclidean version of \eq{rt20} given by .
\begin{equation}
G_E = \frac{1}{\pi^2} \int_0^\infty d\omega\ (\sinh \pi \omega) \, K_{i\omega} (\mu \rho)\, K_{i\omega}(\mu\rho')\, \mathcal{T}_\omega(\theta-\theta')
\end{equation} 
What is nice is that the thermal factor in the Euclidean sector
  can  also be expressed as a periodic sum in the Euclidean angle; that is, we can easily show that:
 \begin{equation}
 \mathcal{T}_\omega(\theta-\theta') =  \sum_{n=-\infty}^\infty e^{-\omega|\theta -\theta' +2\pi n|}
  \end{equation}
  thereby making the periodicity in the Euclidean, Rindler time obvious. This is yet another hidden thermal feature of the inertial propagator!
This allows us to write the Euclidean, inertial, propagator as a thermal sum:
\begin{equation}
G_E =  \sum_{n=-\infty}^\infty\frac{1}{\pi^2} \int_0^\infty d\omega\ (\sinh \pi \omega) \, K_{i\omega} (\mu \rho)\, K_{i\omega}(\mu\rho')\, e^{-\omega|\theta -\theta' +2\pi n|}
\end{equation} 
This equation has a simple interpretation (which will be explored extensively in \cite{rk-tp}): In the right hand side the $n=0$ terms is just the Euclidean propagator \textit{in the Rindler vacuum}. The periodic, infinite, sum `thermalises' it thereby producing the inertial propagator.

  \end{document}